\documentstyle[times,pramana,epsf,epsfig,floats,amsmath,axodraw]{ias}

\input paperdef

\begin{document}

%\mark{{MSSM}{W boson}{precision observables}}

%%%%%%%%%%%%%%%%%%%%%%%%%%%%%%%%%%%%%%%%%%%%%%%%%%%%%%%%%%%%%%%%%%%%%%%%%%%%%%%
%%%%%%%%%%%%%%%%%%%%%%%%%%%%%%%%%%%%%%%%%%%%%%%%%%%%%%%%%%%%%%%%%%%%%%%%%%%%%%%

\thispagestyle{empty}
\setcounter{page}{0}
\def\thefootnote{\fnsymbol{footnote}}

\begin{flushright}
%DCPT/06/DD\\
%IPPP/06/II\\
MPP--2006--100\\
%PSI--PR--06--PP\\
hep-ph/0611373\\
\end{flushright}

\vspace{1cm}

\begin{center}

{\large\sc {\bf Higher-Order Corrected Higgs Bosons in \fhtt}}
\footnote{two talks given by S.~Heinemeyer at the {\em LCWS06}, 9-13 March
  2006, Bangalore, India}

\vspace{1cm}

{\sc T.~Hahn$^{\,1}$%
\footnote{
email: hahn@feynarts.de
}%
, S.~Heinemeyer$^{\,2}$%
\footnote{
email: Sven.Heinemeyer@cern.ch
}%
, W.~Hollik$^{\,1}$%
\footnote{
email: hollik@mppmu.mpg.de
}%
, H.~Rzehak$^{\,3}$%
\footnote{
email: Heidi.Rzehak@psi.ch
}%
,\\[.5em]  G.~Weiglein$^{\,4}$%
\footnote{
email: Georg.Weiglein@durham.ac.uk
}%
 and K.~Williams$^{\,4}$%
\footnote{
email: k.e.williams@durham.ac.uk
}%
}

\vspace*{1cm}

$^1$Max-Planck-Institut f\"ur Physik (Werner-Heisenberg-Institut),\\ 
F\"ohringer Ring 6, D--80805 Munich, Germany 

\vspace*{0.4cm}

$^2$ Instituto de Fisica de Cantabria (CSIC--UC), Santander,  Spain 

\vspace*{0.4cm}

$^3$Paul Scherrer Institut, W\"urenlingen und Villigen, CH--5232
Villigen PSI, Switzerland

\vspace*{0.4cm}

$^4$IPPP, University of Durham, Durham DH1 3LE, U.K.

\end{center}

\vspace*{0.2cm}

\BC {\bf Abstract} \EC
Large higher-order corrections enter the Higgs boson sector of the
MSSM via Higgs-boson self-energies. 
Their effects have to be taken into account for the correct treatment
of loop-corrected Higgs-boson mass eigenstates as external (on-shell)
or internal particles in Feynman diagrams. 
We review how the loop corrections, including momentum dependence and
imaginary contributions, are correctly taken into account for external
(on-shell) Higgs boson and how effective couplings can be derived. The
proceedures are implemented in the code \fhtt.

\def\thefootnote{\arabic{footnote}}
\setcounter{footnote}{0}

\newpage

%%%%%%%%%%%%%%%%%%%%%%%%%%%%%%%%%%%%%%%%%%%%%%%%%%%%%%%%%%%%%%%%%%%%%%%%%%%%%%%
%%%%%%%%%%%%%%%%%%%%%%%%%%%%%%%%%%%%%%%%%%%%%%%%%%%%%%%%%%%%%%%%%%%%%%%%%%%%%%%

\title{Higher-Order Corrected Higgs Bosons in \fhtt}

\author{T.~Hahn$^1$,
        S.~Heinemeyer$^2$, 
        W.~Hollik$^1$, 
        H.~Rzehak$^3$, 
        G.~Weiglein$^4$,
        K.~Williams$^4$}
\address{$^1$ Max-Planck-Institut f\"ur Physik, 
              F\"ohringer Ring 6, D--80805 Munich, Germany\\
         $^2$Instituto de Fisica de Cantabria (CSIC-UC), Santander, Spain\\
         $^3$Paul Scherrer Institut, W\"urenlingen und 
                          Villigen, CH--5232 Villigen PSI, Switzerland\\
         $^4$ IPPP, University of Durham, Durham DH1 3LE, U.K.}


\pacs{2.0}

\abstract{
Large higher-order corrections enter the Higgs boson sector of the
MSSM via Higgs-boson self-energies. 
Their effects have to be taken into account for the correct treatment
of loop-corrected Higgs-boson mass eigenstates as external (on-shell)
or internal particles in Feynman diagrams. 
We review how the loop corrections, including momentum dependence and
imaginary contributions, are correctly taken into account for external
(on-shell) Higgs boson and how effective couplings can be derived. The
proceedures are implemented in the code \fhtt.

}

\maketitle

%%%%%%%%%%%%%%%%%%%%%%%%%%%%%%%%%%%%%%%%%%%%%%%%%%%%%%%%%%%%%%%%%%%%%%%%%%%%%%%
%%%%%%%%%%%%%%%%%%%%%%%%%%%%%%%%%%%%%%%%%%%%%%%%%%%%%%%%%%%%%%%%%%%%%%%%%%%%%%%

\section{Introduction}

The search for Higgs bosons is a crucial test of
Supersymmetry (SUSY) which can be performed with the present and the
next generation of accelerators. A precise prediction for the
production and decay processes of the Higgs bosons in terms
of the relevant SUSY parameters is necessary in order to determine the
discovery and exclusion potential of the
Tevatron~\cite{D0bounds,CDFbounds,Tevcharged}, 
and for 
physics at the LHC~\cite{atlastdr,cmshiggs,HcoupLHCSM,schumi} and the 
ILC~\cite{tdr,nlc,jlc,lhcilc,Snowmass05Higgs}.
A precise prediction of Higgs-boson properties is also required as
input for (current) electroweak precision analyses, see e.g.\
\citeres{MWweber,ehow} 

Since the soft SUSY-breaking parameters can in general be complex, it
is necessary to take into account the effects of complex phases. 
In the Minimal Supersymmetric Standard Model
with complex parameters (cMSSM) Higgs physics is affected by complex
parameters entering via loop corrections. 
In particular, these are the Higgs mixing parameter, $\mu$, the trilinear
couplings, $A_f$, $f = t, b, \tau, \ldots$, and the gaugino mass parameters
$M_1$, $M_2$, $M_3$, where $|M_3| \equiv m_{\tilde g}$ (the gluino
mass). As a consequence, once loop corrections are taken into account the  
neutral Higgs bosons are no longer ${\cal CP}$-eigenstates, 
but mix with each other~\cite{mhiggsCPXgen},
\begin{equation}
(h, H, A) \to h_1, h_2, h_3~.
\end{equation}
As tree-level input parameters in the Higgs sector (besides the gauge
couplings and the $Z$~boson mass) it is convenient to choose the mass
of the charged Higgs 
boson, $M_{H^\pm}$, and the ratio of the two vacuum expectation values,
$\tan\beta$.

In order to obtain the prediction for the Higgs masses beyond lowest order, 
the poles of the Higgs propagators have to be determined. Since the
propagator poles are located in the complex plane, we define the
physical mass of each particle according to the real part of the complex
pole.

Neglecting the mixing with the Goldstone bosons and the (longitudinal
parts) of the $Z$~boson (see \citere{mhcMSSMlong} for details) one can
write the propagator matrix of 
the neutral Higgs bosons $h, H, A$ as a $3 \times 3$ matrix,
$\De_{hHA}(p^2)$. It is related to the $3 \times 3$ matrix of the irreducible
vertex functions by
\begin{equation}
\De_{hHA}(p^2) = - \left(\hat{\Gamma}_{hHA}(p^2)\right)^{-1} ,
\label{eq:propagator}
\end{equation}
where 
\begin{align}
\label{eq:invprophiggs}
  \hat{\Gamma}_{hHA}(p^2) &= {\rm i} \left[p^2 \id - \matr{M}_{\mathrm{n}}(p^2)
                               \right], \\[.5em]
  \matr{M}_{\mathrm{n}}(p^2) &=
  \begin{pmatrix}
    \mh^2 - \ser{hh}(p^2) & - \ser{hH}(p^2) & - \ser{hA}(p^2) \\
    - \ser{hH}(p^2) & \mH^2 - \ser{HH}(p^2) & - \ser{HA}(p^2) \\
    - \ser{hA}(p^2) & - \ser{HA}(p^2) & \mA^2 - \ser{AA}(p^2)
  \end{pmatrix}. %\notag
\label{eq:Mn}
\end{align}
Inversion of $\hat{\Gamma}_{hHA}(p^2)$ yields for the diagonal Higgs
propagators ($i = h, H, A$)
\begin{equation}
\De_{ii}(p^2) = \frac{i}{p^2 - m_i^2 + \ser{ii}^{\rm eff}(p^2)} ,
\label{eq:higgsprop}
\end{equation}
where $\De_{hh}(p^2)$, $\De_{HH}(p^2)$, $\De_{AA}(p^2)$ are the $(11)$,
$(22)$, $(33)$ elements of the $3 \times 3$ matrix $\De_{hHA}(p^2)$,
respectively. The structure of \refeq{eq:higgsprop} is formally the same
as for the case without mixing, but the usual self-energy is replaced by
the effective quantity $\ser{ii}^{\rm eff}(p^2)$ which contains mixing
contributions of the three Higgs bosons. It reads (no summation over 
$i, j, k$)
\begin{align}
\ser{ii}^{\rm eff}(p^2) &= \ser{ii}(p^2) - {\rm i} 
\frac{2 \hat{\Gamma}_{ij}(p^2) \hat{\Gamma}_{jk}(p^2) \hat{\Gamma}_{ki}(p^2) -
      \hat{\Gamma}^2_{ki}(p^2) \hat{\Gamma}_{jj}(p^2) -
      \hat{\Gamma}^2_{ij}(p^2) \hat{\Gamma}_{kk}(p^2)
     }{\hat{\Gamma}_{jj}(p^2) \hat{\Gamma}_{kk}(p^2) - 
       \hat{\Gamma}^2_{jk}(p^2)
      } ,
\label{eq:sigmaeff}
\end{align}
where the $\hat{\Gamma}_{ij}(p^2)$ are the elements of the $3 \times 3$
matrix $\hat{\Gamma}_{hHA}(p^2)$ as specified in
\refeq{eq:invprophiggs}.

For completeness, we also state the expression for the off-diagonal 
Higgs propagators. It reads ($i \neq j$, no summation over $i, j, k$)
\begin{align}
\De_{ij}(p^2) = \frac{\hat{\Gamma}_{ij} \hat{\Gamma}_{kk} -
                      \hat{\Gamma}_{jk} \hat{\Gamma}_{ki}}{
   \hat{\Gamma}_{ii}\hat{\Gamma}_{jj}\hat{\Gamma}_{kk}
   + 2 \hat{\Gamma}_{ij}\hat{\Gamma}_{jk}\hat{\Gamma}_{ki}
   - \hat{\Gamma}_{ii}\hat{\Gamma}_{jk}^2
   - \hat{\Gamma}_{jj}\hat{\Gamma}_{ki}^2
   - \hat{\Gamma}_{kk}\hat{\Gamma}_{ij}^2
                     } ,
\label{eq:higgsprop2}
\end{align}
where we have dropped the argument $p^2$ of the $\hat{\Gamma}_{ij}(p^2)$
appearing on right-hand side for ease of notation.

The complex pole ${\cal M}^2$ of each propagator is determined as 
the solution of
\begin{equation}
{\cal M}_i^2 - m_i^2 + \ser{ii}^{\rm eff}({\cal M}_i^2) = 0 .
\end{equation}
Writing the complex pole as 
\begin{equation}
{\cal M}^2 = M^2 - i M \Ga ,
\end{equation}
where $M$ is the mass of the particle and $\Ga$ its width,
and expanding up to first order in $\Ga$ 
around $M^2$ yields the following
equation for $M_i^2$,
\begin{equation}
M_i^2 - m_i^2 + \re \ser{ii}^{\rm eff}(M_i^2) +
\frac{\im\ser{ii}^{\rm eff}(M_i^2) \, 
      \left(\im\ser{ii}^{\rm eff}\right)^\prime(M_i^2)
      }{1 + \left(\re\ser{ii}^{\rm eff}\right)^\prime(M_i^2)
       } = 0 .
\label{eq:massmaster}
\end{equation}
The short-hand notation $f^{\prime}(p^2) \equiv d \, f(p^2)/(d \, p^2)$
has been used, and $M_i$ denotes the loop-corrected mass,
while $m_i$ is the lowest-order mass ($i = h, H, A$).

%We define the loop-corrected mass eigenvalues according to
%\begin{equation}
%M_{h_1} \leq M_{h_2} \leq M_{h_3} .
%\label{eq:mh123}
%\end{equation}
%In our numerical analysis (and in the code \fhtt) we determine the
%mass eigenvalues using an iterative diagonalization proceedure,
%which avoids the expansion in the width $\Ga$ used for deriving
%\refeq{eq:massmaster}, see \citere{Hahn:2006hr} for further details. 

While the Higgs-boson masses $M_i^2$ can in principle directly be
determined from \refeq{eq:massmaster} by means of an iterative procedure 
(since $M_i^2$ appears as argument of the self-energies in
\refeq{eq:massmaster}), it is often more convenient to determine the
mass eigenvalues from a diagonalization of the mass matrix in 
\refeq{eq:Mn}. 
In the code \fhtt we perform a numerical diagonalization of \refeq{eq:Mn}
using an iterative Jacobi-type algorithm~\cite{Hahn:2006hr}. 
The mass eigenvalues $M_i$ are then determined as the zeros of the
function $\mu^2_i(p^2) - p^2$, where $\mu^2_i(p^2)$ is the $i$th eigenvalue 
of the mass matrix in \refeq{eq:Mn} evaluated at $p^2$.
Insertion of the resulting eigenvalues $M_i$ into \refeq{eq:massmaster} 
verifies (to \order{\Ga}) that each eigenvalue indeed corresponds to
the appropriate (complex pole) solution of the propagator.
We define the loop-corrected mass eigenvalues according to
\begin{equation}
M_{h_1} \leq M_{h_2} \leq M_{h_3} .
\label{eq:mh123}
\end{equation}

%\medskip
In our determination of the Higgs-boson masses we take into account all
imaginary parts of the Higgs-boson self-energies 
(besides the term with imaginary parts 
appearing explicitly in \refeq{eq:massmaster}, there are also products
of imaginary parts in $\re \ser{ii}^{\rm eff}(M_i^2)$).
The effects of the imaginary parts of the Higgs-boson self-energies on
Higgs phenomenology can be 
especially relevant if the masses are close to each other. 
This has been analyzed in \citere{imagSE1} taking into account the
mixing between the two heavy neutral Higgs bosons, where the complex
mass matrix has been diagonalized with a complex mixing angle,
resulting in a non-unitary mixing matrix.
The effects of imaginary parts of the Higgs-boson self-energies on
physical processes with s-channel resonating Higgs bosons
are discussed in \citeres{imagSE1,imagSE2,imagSE3}. In
\citere{imagSE1} only the one-loop corrections from the
$t/\Stop$~sector have been taken into account for the $H$--$A$~mixing,
analyzing the effects on resonant Higgs production at a photon collider.
In \citere{imagSE2} the full one-loop imaginary parts of the
self-energies have been evaluated for the mixing of the three neutral
MSSM Higgs bosons. The effects have been analyzed for resonant Higgs
production at the LHC, the ILC and a photon collider (however, the
corresponding effects on the Higgs-boson masses have been neglected). 
In \citere{imagSE3} the $\Stop/\Sbot$ one-loop contributions (neglecting the
$t/b$ corrections) on the $H$--$A$ mixing for resonant Higgs
production at a muon collider have been discussed.
Our calculation incorporates for the first time the complete effects
arising from the imaginary parts of the one-loop self-energies in the 
neutral Higgs-boson propagator
matrix, including their effects on the Higgs masses and the Higgs
couplings in a consistent way.

As described above, the solution for the Higgs-boson masses in the
general case where the full momentum dependence and all imaginary parts
of the Higgs-boson self-energies are taken into account is numerically
quite involved. 
It is therefore of interest to consider also approximate methods 
for determining the Higgs-boson masses.
Instead of keeping the full momentum dependence in
\refeq{eq:Mn}, the ``$p^2$ on-shell''
approximation consists of setting the arguments of
the self-energies appearing in \refeq{eq:Mn} to the tree-level masses
according to ($i,j = h,H,A$)
\BEA
\label{p2onshell}
\mbox{$p^2$ on-shell approximation: } \quad
\hSi_{ii}(p^2) &\to& \hSi_{ii}(m_i^2) \\
\hSi_{ij}(p^2) &\to& \hSi_{ij}((m_i^2 + m_j^2)/2) ~. \non 
\EEA
In this way the Higgs-boson masses can simply be obtained as the
eigenvalues of the (momentum-inpendent) matrix of \refeq{eq:Mn}. 
The ``$p^2$ on-shell'' approximation has the benefit that it 
removes all residual dependencies on the field renormalization 
constants that cannot be avoided in the iterative procedure.

Instead of setting the momentum argument of the renormalized
self-energies to the tree-level masses, in the ``$p^2 = 0$''
approximation the momentum dependence of the self-energies is neglected
completely ($i,j = h,H,A$),
\BEA
\label{eq:p20approx}
\mbox{$p^2 = 0$ approximation: } \quad
\hSi_{ii}(p^2) &\to& \hSi_{ii}(0) \\
\hSi_{ij}(p^2) &\to& \hSi_{ij}(0) ~. \non 
\EEA
In the ``$p^2 = 0$'' approximation the masses are identified with 
the eigenvalues of $\matr{M}_{\mathrm{n}}(0)$ (see \refeq{eq:Mn})
instead of the true pole masses. This approximation is
mainly useful for comparisons with effective-potential calculations.
The matrix $\matr{M}_{\mathrm{n}}(0)$ is hermitian by construction.

%%%%%%%%%%%%%%%%%%%%%%%%%%%%%%%%%%%%%%%%%%%%%%%%%%%%%%%%%%%%%%%%%%%%%%%%%%%%%%%
%%%%%%%%%%%%%%%%%%%%%%%%%%%%%%%%%%%%%%%%%%%%%%%%%%%%%%%%%%%%%%%%%%%%%%%%%%%%%%%

\section{Amplitudes with external Higgs Bosons}
\label{sec:extHiggs}

In evaluating processes with external (on-shell) Higgs bosons beyond 
lowest order one has to account for the mixing between the Higgs
bosons in order to ensure that the outgoing particle has the correct
on-shell properties such that the S~matrix is properly normalized. 
This gives rise to finite wave-function normalization factors.%
\footnote{The introduction of these factors can in principle be avoided 
by using a renormalization scheme where all involved particles obey on-shell 
conditions from the start, but it is often more convenient to work in a
different scheme like the \drbar\ scheme for the field
renormalizations~\cite{HiggsDRbar}.} 
For the case of $2 \times 2$ mixing appearing in the MSSM with real
parameters (rMSSM) for the
mixing between the two neutral $\cp$-even Higgs bosons $h$ and $H$, 
which is analogous to the mixing of the
photon and $Z$ boson in the Standard Model, the relevant wave function
normalization factors are well-known, see e.g.\
\citeres{mhiggsf1lC,eennH}. An amplitude with an external Higgs boson,
$i$,  receives the corrections ($i,j = h,H$, no summation over $i,j$)
\BE
%i ~:~
\sqrt{\hat Z_i} \KL \Ga_i \; + \; \hat Z_{ij} \Ga_j \KR
                                                   \quad (i\neq j)~,
\label{eq:Vert_i}
\end{equation}
where the
$\Ga_{i,j}$ denote the one-particle irreducible Higgs vertices, and 
\begin{align}
\label{eq:Zi}
\hat Z_i &= \KKL 1 + \re \hSi_{ii}^{\prime}(p^2) - 
     \re \KL \frac{\KL \hSi_{ij}(p^2) \KR^2}
     {p^2 - m_j^2 + \hSi_{jj}(p^2)} 
     \KR^{\prime}~\KKR^{-1}_{\Bigr| p^2 = M_i^2}\,, 
                                                                 \\[.5em]
\label{eq:Zij}
\hat Z_{ij} &= -\frac{\hSi_{ij}(M_i^2)}{M_i^2 - m_j^2 + \hSi_{jj}(M_i^2)}\, .
\end{align}
As before
$m_j$ denotes the tree-level mass, while $M_i$ is the
loop-corrected mass.

In the case of the cMSSM, the formulas above need to be extended to the
case of $3 \times 3$ mixing. A vertex with an external Higgs
boson, $i$,
has the form (with $i,j,k$ all different, $i,j,k = h, H, A$, and no
summation over indices)
\BE
\sqrt{\hat Z_i} \KL \Ga_i \; + \; 
          \hat Z_{ij} \Ga_j \; + \; \hat Z_{ik} \Ga_k + \ldots \KR ~,
\label{eq:zfactors}
\end{equation}
where the ellipsis represents contributions from the mixing with the
Goldstone boson and the $Z$~boson, see \citere{mhcMSSMlong} for more details.
The finite $Z$~factors are given by
\begin{align}
\hat Z_i &= \frac{1}{1 + 
              \left(\re\ser{ii}^{\rm eff}\right)^\prime(M_i^2)} , 
              \label{eq:zi} \\[.5em]
\hat Z_{ij} &= \frac{\De_{ij}(p^2)}{\De_{ii}(p^2)}_{~\Bigr| p^2 = M_i^2} \non \\
            &= \frac{\ser{ij}(M_i^2) 
               \left(M_i^2 - m_k^2 + \ser{kk}(M_i^2)\right) -
               \ser{jk}(M_i^2)\ser{ki}(M_i^2)
                    }{
               \ser{jk}^2(M_i^2) - 
               \left(M_i^2 - m_j^2 + \ser{jj}(M_i^2)\right)
               \left(M_i^2 - m_k^2 + \ser{kk}(M_i^2)\right)
                     } ,
               \label{eq:zij}
\end{align}
where the propagators $\De_{ii}(p^2)$, $\De_{ij}(p^2)$ 
have been given in \refeqs{eq:higgsprop} and (\ref{eq:higgsprop2}),
respectively. Using
\refeq{eq:zfactors} with $Z$ factors specified in \refeqs{eq:zi},
(\ref{eq:zij}) and adding to this expression the mixing contributions of
the Higgs bosons with the Goldstone bosons and the gauge bosons (see 
\citere{mhcMSSMlong} for more details) 
yields the correct normalization of the outgoing Higgs
bosons in the S~matrix.

For convenience we define a matrix $\matr{\tilde Z}_{\mathrm{n}}$
based on the wave function normalization factors. Its elements are given
by (with $\hat Z_{ii} = 1$, $i, j = h, H, A$, and no summation over $i$)
\BE
(\matr{\tilde Z}_{\mathrm{n}})_{ij} := \sqrt{\hat Z_i} \; \hat Z_{ij}~.
\label{eq:defZ}
\end{equation}
Performing a re-ordering of the lines of $\matr{\tilde  Z}_{\mathrm{n}}$ 
such that they correspond to the mass ordering of \refeq{eq:mh123}
results in the matrix $\matr{Z}_{\mathrm{n}}$. 
A vertex with an external (on-shell) Higgs boson $h_i$ is then given by
\BE
(\matr{Z}_{\mathrm{n}})_{i1} \Ga_h +
(\matr{Z}_{\mathrm{n}})_{i2} \Ga_H +
(\matr{Z}_{\mathrm{n}})_{i3} \Ga_A + \ldots
\label{eq:zfactors123}
\end{equation}
where the ellipsis again represents contributions from the mixing with the
Goldstone boson and the $Z$~boson.

%%%%%%%%%%%%%%%%%%%%%%%%%%%%%%%%%%%%%%%%%%%%%%%%%%%%%%%%%%%%%%%%%%%%%%%%%%%%%%%
%%%%%%%%%%%%%%%%%%%%%%%%%%%%%%%%%%%%%%%%%%%%%%%%%%%%%%%%%%%%%%%%%%%%%%%%%%%%%%%

\section{Effective couplings}
\label{sec:intHiggs}

In a general amplitude with internal Higgs bosons, the structure describing 
the Higgs part is given by
$\sum_{ij} \Ga_i \; \De_{ij} \; \Ga_j$,
where the $\Ga_{i,j}$ are as above the one-particle irreducible Higgs
vertices, and the propagators $\De_{ij}$ are given in \refeqs{eq:higgsprop}
and (\ref{eq:higgsprop2}). 
For phenomenological analyses it is often convenient to use approximations of
improved-Born type with effective couplings incorporating leading
higher-order effects. There is no unique prescription how to define such
effective coupling terms. One possibility would be to consider the matrix
$\matr{Z}_{\mathrm{n}}$, defined through
\refeqs{eq:defZ}--(\ref{eq:zfactors123}), as mixing matrix. 
The elements of the matrix $\matr{Z}_{\mathrm{n}}$, however, are in general
complex, so that the $\matr{Z}_{\mathrm{n}}$ is a non-unitary
matrix. Therefore it cannot be interpreted as a rotation matrix. 
If one wants to introduce effective couplings by means of a (unitary) rotation
matrix, it is necessary to make further approximations. 

A possible choice leading to a unitary rotation matrix is the ``$p^2 = 0$''
approximation, which is used in the effective potential approach.
As before, we first consider the case of $2 \times 2$ mixing relevant
for the rMSSM. 
In the ``$p^2 =0$'' approximation defined in \refeq{eq:p20approx}
the momentum dependence in the renormalized self-energies is set to the 
respective tree-level Higgs boson masses, 
so that the derivative in \refeq{eq:Zi} acts
only on the $p^2$~term in the propagator factor. In this limit $\hat Z_i$ 
simplifies to~\cite{hff,eehZhA}
\BE
\mbox{$p^2 = 0$ approximation, $2 \times 2$ mixing: } \quad
\hat Z_i = \ed{1 + \hat Z_{ij}^2}~.
\label{eq:zi22p20}
\end{equation}
For the mixing between the neutral $\cp$-even Higgs bosons $h, H$ this
yields $\hat Z_h = \hat Z_H = \cos^2\De\al$. This corresponds to an
effective coupling approximation where the 
tree-level mixing angle $\alpha$
appearing in the couplings of the neutral $\cp$-even Higgs bosons is
replaced by $\aeff = \al + \De\al$~\cite{hff,eehZhA}.

It is easy to verify that for the $3 \times 3$ mixing case \refeq{eq:zi} 
in the ``$p^2 = 0$'' approximation simplifies to 
\BE
\mbox{$p^2 = 0$ approximation, $3 \times 3$ mixing: } \quad
\hat Z_i = \ed{1 + \hat Z_{ij}^2 + \hat Z_{ik}^2} ~,
\label{eq:zi33p20}
\end{equation}
as a direct generalization of \refeq{eq:zi22p20}.

The matrix $\matr{Z}_{\mathrm{n}}$ defined through
\refeqs{eq:defZ}--(\ref{eq:zfactors123})
goes over into a unitary matrix $\matr{R}_{\mathrm{n}}$ 
in this approximation,
\BE
\mbox{$p^2 = 0$ approximation, $3 \times 3$ mixing: } \quad
\matr{Z}_{\mathrm{n}} \to \matr{R}_{\mathrm{n}}, \quad
\matr{R}_{\mathrm{n}} =
  \begin{pmatrix}
    R_{11} & R_{12} & R_{13} \\
    R_{21} & R_{22} & R_{23} \\
    R_{31} & R_{32} & R_{33}
  \end{pmatrix} .
\end{equation}
The matrix $\matr{R}_{\mathrm{n}}$ diagonalizes the matrix 
$\matr{M}_{\mathrm{n}}(0)$ arising from \refeq{eq:Mn} in the 
``$p^2 = 0$'' approximation. 
$\matr{R}_{\mathrm{n}}$ can therefore be used to connect 
the mass eigenstates $h_1, h_2, h_3$
with the original states $h, H, A$,
\begin{align}
  \begin{pmatrix} h_1 \\ h_2 \\ h_3 \end{pmatrix}_{p^2=0} &\hspace{-0.5em}=
  \matr{R}_{\mathrm{n}} \cdot \begin{pmatrix} h \\ H \\ A
  \end{pmatrix}, \; 
  \matr{R}_{\mathrm{n}} \, \matr{M}_{\mathrm{n}}(0) \,
  \matr{R}_{\mathrm{n}}^\dagger =
  \begin{pmatrix}
    M_{h_1\,p^2=0}^2 & 0 & 0 \\ 
    0 & M_{h_2\,p^2=0}^2 & 0 \\ 
    0 & 0 & M_{h_3\,p^2=0}^2
  \end{pmatrix} .
\label{eq:defR}
\end{align}
Since $\matr{M}_{\mathrm{n}}$ is hermitian, the matrix
$\matr{R}_{\mathrm{n}}$ is unitary.

As shown in \citere{mhcMSSMlong}, a better approximation of the
full result can be achieved by defining the effective couplings in the 
``$p^2$ on-shell'' approximation. 
The unitary matrix $\matr{U}_{\mathrm{n}}$ is defined such that it
diagonalizes the matrix  $\re\KL\matr{M}_{\mathrm{n}}(p^2 \mbox{ on-shell})\KR$ 
arising from \refeq{eq:Mn} in the 
``$p^2$ on-shell'' approximation and neglecting the imaginary parts.
This yields
\begin{align}
  &\mbox{$p^2$~on-shell approx.: } \;
  \begin{pmatrix} h_1 \\ h_2 \\ h_3 \end{pmatrix}_{p^2 {\rm ~on-shell}} 
  \hspace{-0.5em}=
  \matr{U}_{\mathrm{n}} \cdot \begin{pmatrix} h \\ H \\ A
  \end{pmatrix}, \;
\matr{U}_{\mathrm{n}} =
  \begin{pmatrix}
    U_{11} & U_{12} & U_{13} \\
    U_{21} & U_{22} & U_{23} \\
    U_{31} & U_{32} & U_{33}
  \end{pmatrix} , \\[.5em]
  &\matr{U}_{\mathrm{n}} \, \re\KL\matr{M}_{\mathrm{n}}(p^2 \mbox{ on-shell})\KR \,
  \matr{U}_{\mathrm{n}}^\dagger =
  \begin{pmatrix}
    M_{\He,p^2 {\rm ~on-shell}}^2 & 0 & 0 \\ 
    0 & M_{\Hz,p^2 {\rm ~on-shell}}^2 & 0 \\ 
    0 & 0 & M_{\Hd,p^2 {\rm ~on-shell}}^2
  \end{pmatrix} . \non
\label{eq:defU}
\end{align}

The elements of $\matr{U}_{\mathrm{n}}$ can be used to quantify the
extent of $\cp$-violation. 
For example, $U_{13}^2$ can be understood as the $\cp$-odd part in $h_1$,
while $U_{11}^2+U_{12}^2$ make up the $\cp$-even part. 
The unitarity of $\matr{U}_{\mathrm{n}}$ ensures that both parts
add up to 1. 

The elements of $\matr{U}_{\mathrm{n}}$ can
be interpreted as effective couplings of internal Higgs bosons in a
loop diagram, which take into account leading higher-order
contributions.
As an example we show the couplings of a neutral Higgs boson to two gauge
bosons, $VV = ZZ, W^+W^-$. 
Beyond the lowest order in the cMSSM all three neutral Higgs bosons have
a $\cp$-even component, so that all three Higgs bosons have
non-vanishing couplings to two gauge bosons. 
The couplings normalized
to the SM values are given by
\begin{align}
g_{h_i VV} &= U_{i 1} \Sba + U_{i 2} \Cba .
\end{align}

%%%%%%%%%%%%%%%%%%%%%%%%%%%%%%%%%%%%%%%%%%%%%%%%%%%%%%%%%%%%%%%%%%%%%%%%%%%%%%%
%%%%%%%%%%%%%%%%%%%%%%%%%%%%%%%%%%%%%%%%%%%%%%%%%%%%%%%%%%%%%%%%%%%%%%%%%%%%%%%

\section{Conclusions}

Large higher-order corrections enter the Higgs boson sector of the
MSSM via Higgs-boson self-energies. 
Their effects have to be taken into account for the correct treatment
of loop-corrected Higgs-boson mass eigenstates as external (on-shell)
or internal particles in loop diagrams. 
We have shown how the loop corrections can be taken into account in
the treatment of external (on-shell) Higgs bosons, and how effective
couplings, taking into account leading higher-order effects, can be
constructed.  

The proceedure presented here is implemented in the code 
\fhtt \cite{mhcMSSMlong,mhiggslong,mhiggsAEC,feynhiggs}. 
For the Higgs-boson self-energies the complete one-loop
corrections \cite{mhcMSSMlong,mhiggsCPXFD1,mhiggsCPXFDproc,habilSH}
are supplemented by the available
two-loop corrections in the Feynman-diagrammatic approach for the 
MSSM with real parameters~\cite{mhiggslong,mhiggsAEC,mhiggsletter} and
a resummation of the leading (s)bottom 
corrections for complex parameters~\cite{deltamb2}.
Higgs boson decays and production cross sections~\cite{Tev4LHCsigmaH} 
are evaluated using
$\matr{Z}_{\mathrm{n}}$~(\ref{eq:defZ},\ref{eq:zfactors123}), 
ensuring the on-shell
properties of the external particles. Effective Higgs boson couplings are
obtained with the help of $\matr{U}_{\mathrm{n}}$~(\ref{eq:defU}).
In this way {\tt FeynHiggs} provides a precise prediction of Higgs
boson production and decay properties and the respective couplings.

%%%%%%%%%%%%%%%%%%%%%%%%%%%%%%%%%%%%%%%%%%%%%%%%%%%%%%%%%%%%%%%%%%%%%%%%%%%%%%%
%%%%%%%%%%%%%%%%%%%%%%%%%%%%%%%%%%%%%%%%%%%%%%%%%%%%%%%%%%%%%%%%%%%%%%%%%%%%%%%

\end{document}